# Polarization induced instabilities in external four-mirror Fabry-Perot cavities

[1]F. Zomer, [1]Y. Fedala, [2]N. Pavloff, [1]V. Soskov and [1]A. Variola

[1]Laboratoire de l'Accélérateur Linéaire, C.N.R.S./IN2P3 Université Paris sud, Bâtiment 200, BP 34 Orsay Cedex France

2 Laboratoire de Physique Théorique et Modèles Statistiques, Université Paris Sud, Bâtiment 100, F-91405 Orsay Cedex, France

[*]Corresponding author: zomer@lal.in2p3.fr

Abstract. Various four-mirror optical resonators are studied in the perspective of realizing passive stacking cavities. A comparative study of the mechanical stability is provided. The polarization properties of the cavity eigenmodes are described and it is shown that the effect of mirror misalignments (or motions) induces polarization and stacking power instabilities. These instabilities increase with the finesse of the Fabry-Perot cavity. A tetrahedral configuration of the four mirrors is found to minimize the consequences of the mirrors' motion and misalignment by reducing the instability parameter by at least two orders of magnitude.

*OCIS codes:* 000.0000, 999.9999.

# Introduction

Externals Fabry-Perot cavities [1] operated in pulsed regime [2] are considered as promising potential tools for producing high flux monochromatic X or gamma rays from laser-electron beams Compton interaction [3,4,5]. The domain of application of monochromatic X/γ-ray sources is extremely broad, including medical imagery [6], medical radiotherapy [7], coronary angiography [8], material science applied to art craft expertise [9], nuclear radioactive wastes management [10] and High Energy Physics [11]. Various experimental programmes are starting using different electron accelerator technologies and Fabry-Perot cavity geometries (e.g. see [12]).

The high X/gamma-ray flux required for the above-mentioned applications imposes strong constraints on the external optical resonator. In particular, the typical resonator round trip optical path should be of the order of a few meters, whereas the laser beam waist inside the cavity must be of the order of a few tens of microns. This means that two-mirror cavities should be discarded since such a small beam waist corresponds to a highly unstable concentric configuration [1]. One is thus lead to choose four-mirror cavities (bow-tie or Z folded) which are known to provide stable operation conditions even when the cavity mode waist is small. However, four-mirror cavities have a drawback which is related to the fact that the circulating light beam is reflected on the cavity mirrors under non vanishing incidence. More precisely, the high reflection mirror coatings needed for reaching a high finesse are made of quarter wave stacks [13] and we will show that under non normal-incidence, small mirror misalignments and motion induce significant fluctuations of the cavity's eigenmode polarization. Since a variation of polarization coupling leads, in turn, to a variation of the laser beam power stored inside the cavity, it is of fundamental importance to study and to quantify the effects of the unavoidable mirror motion or residual misalignment, especially for high finesse cavities. This is the purpose of the present article.

Previous studies of polarization instability in Fabry-Perot cavities exist which consider situations where a non linear coupling is induced by a material located inside the cavity [14]. In our work, we have only considered the instability induced by coupling the incident light polarization vector, which is assumed to be fixed, with the eigenmode of an empty cavity whose polarization is varying because of the mirror motion or misalignment. To our knowledge, geometrically induced polarization instabilities, such as those we are interested in, have never been studied in linear passive cavities.

This article is organized as follows: the typical applications we have in mind and the specific constrains associated with them (high flux, high enhancement factor, high stability …) are presented in section 1. The formalism used for computing the transport of polarization inside planar and non planar four-mirror cavities is described in section 2. The geometries of planar and non planar cavities are also introduced in this section. Numerical results are presented in section 3.

## Section 1 Laser electron interaction and technical constraints on the four-mirror cavity geometry

Ultrarelativistic particles beams are often exploited as radiation sources due to the attractive characteristics of the emitted photons. Depending on the emission mechanism different energy ranges and brightness can be achieved. At present a lot of Synchrotron light sources are operational in the world. Synchrotron emission is a characteristic of charged particles bended in curved trajectories and, due to the significant flux produced, it is used in a wide range of applications [15].

Among the other various light source mechanisms, the Compton laser-electron beams scattering was proposed in [3] but was not considered as photon source due its very small cross-section. But recent improvements in lasers, accelerators and optical resonators allow to have at one's disposal high density electron bunches and high energy photon pulses.

In Compton scattering the photons are produced by the kinematical collision between a charged particle and a photon. In experimental terms this implies to collide a charged electron bunch with a laser pulse (see Figure 1) in an interaction point (IP). The relation between the scattered photon emission angle and its energy is univocal. In the ultrarelativistic limit, for head-on collisions and in the laboratory rest frame, this relation reads as:

$$E_f(\theta) \approx E_i \frac{4\gamma^2}{1+\gamma\theta^2}$$

Where $E_f$ and $E_i$ are respectively the photon energy after and before the collision and $\theta$ is the diffusion angle, *i.e.* the angle between the outgoing photon and the incident electron beam. The incident photons energy is thus boosted by a factor $4\gamma^2$, where $\gamma = E_e/m_e c^2$ is the Lorentz relativistic factor and $E_e$ and $m_e$ are electron energy and mass. In Figure 2(a) $E_f(\theta)/E_f(0)$ is plotted as a function of the diffusion angle for three different electron beam energies $E_e$ = 5 MeV ($\gamma$ =10), $E_e$ = 50 MeV ($\gamma$ = 100), $E_e$ = 150 MeV ($\gamma$ =300) and for a laser beam wavelength $\lambda$=1µm. In Figure 2(b) the scattered photon energy is shown as a function of θ for $E_i$=1eV and 2 eV (that is for the three laser beam wavelengths $\lambda$ 1µm, 0.5µm). From these figures, one can underline the angular dependence of the backscattered photons energy, i.e. the 1/γ emission opening angle typical of the relativistic electron radiations. This allows selecting a spectral width with a simple diaphragm system.

In the light of these considerations, it is possible to summarize the attractive characteristics of the Compton scattering, the photon energy boost, the angular-energy dependence and the directivity. The first one leads to the production of hard X-rays or even 1.6-160 keV gamma rays with relatively low electron beam energies (Ee ~10 MeV-100 MeV for $E_i$=1eV) thus reducing the costs of the experimental apparatus. The second one allows the flux monochromatization (up to few percent of the emitted spectrum) only by selecting part of the angular spectrum (i.e. by diaphragming) and the third one provides a high brilliance photon flux in the direction of the impinging electron beam.

Maximizing the average flux is crucial for the main applications of the Compton scattering. Here we distinguish between two energy ranges:

In the low emitted photon energies domain (10-100keV), important developments of the Compton associated technologies are expected to create a generation of high flux ( $10^{11}$-$10^{13}$ ph/sec), quasi-monochromatic ( ΔE/E=1-10%), low beam divergence (few mrad or less), low cost (few M$) and compact (few meters in circumference electron ring) radiation machines. These characteristics can be exploited in a large variety of research domains as described in the introduction. However, to achieve the required flux (comparable to that of the first or second generation synchrotron sources) with a device that can be easily installed in an hospital, an university or a museum laboratory it is indispensable to bring together the best performances of both electron accelerators and laser systems. These projects require high quality electron beams, e.g. for a small storage ring of few meters of diameter we can consider an electron bunch charge of a 0.1-1 nC, a bunch length of 5 ps and a high focusing system to reach beam sizes of the order of ten micron or so in IP. Targeting an X-ray flux of $10^{13}$ ph/sec, as required by radiotherapy medical application, the requirements for the optical system are: a laser beam of ~100W average

power (repetition frequency of ~100MHz, 1 ps pulse length and wavelength λ 1 µm), a laser pulse staking inside a passive Fabry Perot resonator with a power enhancement factor of 10000 in order to reach 1 MW average power at the IP.

High energies photons produced by electron - laser collisions are envisaged [11] to generate polarized positrons $e^+$ by $e^+$- e- conversion of the produced gamma rays in amorphous targets. In this case, higher energy electron beams (1-2 GeV for λ=1 µm) are required as well as a circularly polarized laser beam. The final degree of polarization of the positron beam depends crucially on the laser degree of circular polarization and its reliability and stability are essential to ensure the performances of the $e^+$ source. The laser beam waist must be reduced to few tenths of microns and the megawatt average power inside the cavity is also required.

In summary, the constraints imposed on the optical cavity design by the requested X-rays and gamma-ray fluxes are the following: good intra-cavity stacking power stability below the percent level; good stability of the degree of circular polarization (for high energy gamma ray application) also better than the percent level; small laser beam waist at the IP, from ~20µm to ~100µm. As shown in Figure 1, the design of the cavity must also include the electron beam pipe whose diameter is usually of the order of few centimetres and last but not least the distance between the spherical mirrors must be long enough (typically around 1m) in order to reduce the laser-electron beams crossing angle. This means that small ultra stable monolithic resonators design cannot be used here and that tilting actuators must be integrated in the mirror mounts to align the cavity. Therefore a cavity with a weak sensitivity to the vibrations induced by the noisy environment of an electron accelerator should be designed.

## Section 2 Formalism

The performances of four-mirror Fabry-Perot cavities of various geometries will be studied. The simpler configurations are planar and are depicted in Figure 3: U-folded (Figure 3.a), Z-folded (Figure 3.b) and bow-tie (Figure 3.c). The non planar extensions of these configurations are shown in Figure 3.d (U-folded), Figure 3.e (Z-folded) and Figure 3.f (bow-tie). For the state of convenience, a relative angle $\phi$ has been introduced such that $\phi=0$, $\pi$ corresponds to the planar geometries and $\phi=\pi/2$ to the 'tetrahedron' and the non planar U folded (a six-mirror version of which was used in [16]) and Z folded cavities. The 'tetrahedron cavity' explicitly is shown in Figure 4. In this case, the reflections on mirror 1 and 3 are located on axis $Ox$, symmetrically to the plane $yOz$, and the reflections on mirrors 2 and 4 are located on axis $Oy$, symmetrically to the plane $xOz$ (see Figure 4).

In Figure 3 and Figure 4, mirrors 1 and 2 are flat and mirrors 3 and 4 are spherical. The radii of curvatures of the spherical mirrors will be chosen in order to minimize the waists of the cavity modes for each geometrical configuration. The reference frame $x,y,z$ is also shown in Figure 3 and Figure 4 together with the length parameters $L$, $h$ and $d$ and the incident angle $\theta_0$. For the sake of simplicity, the cavity geometries are constructed in such a way that the angle of incidence $\theta_0$ is the same on all mirrors, but the formalism described below can handle any other configurations.

The cavities are said to be perfectly aligned when the mirror geometrical centers are located at the reference points $M_{Ci}=(X_{0i},Y_{0i},Z_{0i})$ with $i=1,..4$. The normal vectors at these points are denoted by $\mathbf{N_1}$, $\mathbf{N_2}$ and $\mathbf{N_{03}}$, $\mathbf{N_{04}}$ for the two flat and the two spherical mirrors (the subscripts 03 and 04 indicate that the normal vectors are taken at the geometrical centre of mirrors 3 and 4 respectively). The reflection points on the mirrors are denoted by $M_i$ (with $i=1,..,4$) and $M_i=M_{Ci}$ when the cavity is aligned (see Figure 4).

The misalignment of mirror $i$ is described by five parameters: $\Delta X_i$, $\Delta Y_i$, $\Delta Z_i$, $\Delta\theta_{xi}$ and $\Delta\theta_{yi}$ which characterize the departure in position and angle from perfect alignment. Precisely, the geometrical centers of the misaligned mirrors and their normal directions at these points read as

$$M_{Ci} = (X_i, Y_i, Z_i) = (X_{0i} + \Delta X_i, Y_{0i} + \Delta Y_i, Z_{0i} + \Delta Z_i), \quad i=1,2,3,4$$
$$\mathbf{n_i} = \Re_x(\Delta\theta_{ix})\Re_y(\Delta\theta_{iy})\mathbf{N_i}, \quad i=1,2,03,04$$

where $\Re_\alpha(\Delta\theta_{\alpha j})$ is the 3 dimension matrix describing the rotation of an angle $\Delta\theta_{\alpha j}$ around the axis $\alpha$ ($\alpha=x$ or $y$) in $\mathbb{R}^3$.

## *Accurate optical axis calculation of misaligned cavities*

Given a set of misalignment parameters $\Delta X_i$, $\Delta Y_i$, $\Delta Z_i$, $\Delta\theta_{xi}$ and $\Delta\theta_{yi}$, the method usually used to determine the optical axis of a slightly misaligned complex cavity is the extended ABCD matrix formalism [17,18]. However, we need here to accurately determine the angle of incidence on the cavity mirrors in order to adequately compute the reflection coefficient of the multilayer coatings. Yet we found more suitable to use Fermat's principle [19] which embodies the exact mirror shapes and which can be simply implemented iteratively on the basis of Newton-Rafstone algorithm allowing to reach very high numerical accuracy. The approach is the following: we start by expressing the surface equations of the misaligned mirrors $z=f_i(x,y)$, $i=1..4$. Then, we arbitrarily choose points $M_i=(x_i,y_i,z_i)$ on the misaligned mirrors (with $z_i=f_i(x_i,y_i)$) and evaluate the closed orbit corresponding to the round trip optical path

$$\Lambda = \left\|\overrightarrow{M_2M_1}\right\| + \left\|\overrightarrow{M_3M_2}\right\| + \left\|\overrightarrow{M_4M_3}\right\| + \left\|\overrightarrow{M_1M_4}\right\|$$

for planar and non planar bow-tie cavities. For planar and non planar U or Z folded type cavities the round trip path corresponds to six reflections and the previous expression generalizes to:

$$\Lambda = \|\overrightarrow{M_3 M_1}\| + \|\overrightarrow{M_4 M_3}\| + \|\overrightarrow{M_2 M_4}\| + \|\overrightarrow{M_5 M_4}\| + \|\overrightarrow{M_6 M_5}\| + \|\overrightarrow{M_1 M_6}\| \tag{0.1}$$

where points $M_5$ and $M_6$ are respectively located on mirrors 4 and 3. It turns out that in all our simulations the physical solution which minimizes $\Lambda$ in eq. (0.1) (for U or Z folded cavities) always corresponds to a self-retracing orbit with $M_5=M_4$ and $M_6=M_3$, though we do not impose any a priori condition to the optical path except that the reflection points should lie on the mirror's surfaces.

According to Fermat's principle, the physical trajectories determining the optical axis correspond to minima of $\Lambda$; therefore the coordinates of the actual reflection points on the mirrors are given by the solution of the equations $\{\partial \Lambda / \partial x_i = 0, \partial \Lambda / \partial y_i = 0\}_{i=1,\ldots,4}$. Since these equations are non linear in $x_i$ and $y_i$, in a first stage we perform a first order expansion in $(X_i - x_i)$, $(Y_i - y_i)$ resulting in a system of linear equations which is solved numerically (eight equations for bow-tie cavities and twelve for the U and Z folded types) using the Matlab software [20]. Once the unknown coordinates $x_i$ and $y_i$ are determined accordingly, we reconstruct the optical path using the law of reflections on the mirrors without any approximation starting from the direction $\overrightarrow{M_3 M_4}$. After a round trip, the point of arrival on mirror 3 is denoted by $M_3'$ and the distance $\|\overrightarrow{M_3 M_3'}\|$ is computed. Finally, one substitutes $X_i \to x_i$ and $Y_i \to y_i$ in order to iterate the procedure until the numerical precision is reached such that $\|\overrightarrow{M_3 M_3'}\| \leq 10^{-12}$ mm.

We checked that this method gives results in good agreement with the extended ABCD matrix formalism. However, in order to precisely check our numerical results, we have also computed the optical axis for planar cavity with planar misalignments using a simple independent non-iterative method. For the sake of clarity, the method is only described here for the bow-tie geometry. This method is based on the fact that mirror reflections are isometries and that two successive reflections are equivalent to the product of a space rotation and a translation [21]. Neglecting the translation, the reflections on the flat mirrors 1 and 2 are equivalent to a rotation of an angle $\cos\alpha_{12} = \mathbf{n_1} \cdot \mathbf{n_2} / 2$ around the direction $\mathbf{n_1} \times \mathbf{n_2}$. Since the optical axis is restricted in this case to lie in the plan of the cavity, one can easily write the condition for a ray direction to reproduce itself after a round trip. This leads to the following condition for the rotation matrix describing the reflections on mirrors 3 and 4

$$\Re_{\mathbf{n_3} \times \mathbf{n_4}}(\alpha_{34}) = \Re^{-1}_{\mathbf{n_1} \times \mathbf{n_2}}(\alpha_{12}) \Rightarrow \mathbf{n_3} = -\Re_{\perp}(\delta\alpha)\mathbf{n_2} \text{ if } \mathbf{n_4} = -\Re_{\perp}(-\delta\alpha)\mathbf{n_1}$$

where $\delta\alpha$ is the only free parameter. Since unique points $M_3$ and $M_4$ correspond to the normal vectors $\mathbf{n_3}$ and $\mathbf{n_4}$, $\delta\alpha$ is determined numerically by matching the point of departure and the return point after a round trip on the surface of mirror 3.

The comparison between the results obtained with this method (in the planar case) and the iterative Fermat's method shows a perfect agreement within the Matlab software numerical precision. It should be mentioned that, since the product of two rotations in tree dimensional space obeys to complex Quaternion algebra, we didn't find any way to efficiently extend this simple method to non planar configurations.

*Jones round-trip matrix*

We first concentrate on the bow-tie and tetrahedron cavities. The calculation of the Jones round trip matrix of a non planar oscillator has been described in [22]. For each misalignment configuration, once the optical axis has been determined, the incidence angles $\theta_i$ on the cavity mirrors are obtained. These incidence angles may differ from the nominal incidence angle $\theta_0$ that we assume having been used to define the thickness on the coating layers. Let us denote by $r_i$ the reflection matrix of mirror $i$ in the $\{\mathbf{s_i},\mathbf{p_i}\}$ basis attached to the plane of incidence

$$r_i = \begin{pmatrix} \rho_{is} \exp(i\varphi_{is}) & 0 \\ 0 & \rho_{ip} \exp(i\varphi_{ip}) \end{pmatrix}$$

The real parameters $\rho_{is}$, $\rho_{ip}$, $\varphi_{is}$ and $\varphi_{ip}$ can be computed using the matrix propagation formalism in dielectric multilayers [23]. The mirrors coating that we consider have the following multilayer structure: a $\lambda/2$ $SiO_2$ protection layer, $N$ $\lambda/4$ double layers $Ta_2O_5/SiO_2$, a $\lambda/4$ $Ta_2O_5$ layer and a fused silica substrate. If $\theta_i$ is different from $\theta_0$ for which the coating has been optimize, one has $\varphi_{ip} - \varphi_{is}$ $\pi$. This means that an $s$ and $p$ waves accumulate a different phase after a cavity round trip, *i.e* they will resonate at different frequencies although the cavity is made of an even number of mirrors (see e.g. [24] for the extreme case of an odd number of mirrors).

From the knowledge of the $r_i$'s, the Jones matrix $J$ is obtained by accounting for the change of frame when going from one plane of incidence to another [22,25,26]:

$$J = r_1 N_{41} r_4 N_{34} r_3 N_{23} r_2 N_{12} \tag{0.2}$$

where

$$N_{i,i+1} = \begin{pmatrix} \mathbf{s_i} \cdot \mathbf{s_{i+1}} & \mathbf{p'_i} \cdot \mathbf{s_{i+1}} \\ \mathbf{s_i} \cdot \mathbf{p_{i+1}} & \mathbf{p'_i} \cdot \mathbf{p_{i+1}} \end{pmatrix} \tag{0.3}$$

Denoting by $\mathbf{k_i}$ and $\mathbf{k_{i+1}}$ the incident and reflected wave vectors of mirror $i$ (see Figure 4) and $k_i = |\mathbf{k_i}|$, the vectors $\mathbf{s}$, $\mathbf{p}$ and $\mathbf{p'}$ appearing in eq. (0.3) are given by $\mathbf{s_i} = \mathbf{n_i} \times \mathbf{k_{i+1}}/k_{i+1}$ and $\mathbf{p_i} = \mathbf{k_i} \times \mathbf{s_i}/k_i$, $\mathbf{p_i'} = \mathbf{k_{i+1}} \times \mathbf{s_i}/k_{i+1}$, Note that $J$ in eq. (0.2) is expressed the basis $\{\mathbf{s_1}, \mathbf{p_1}\}$ and that the orthogonal basis $\{\mathbf{s_i}, \mathbf{p_i}, \mathbf{k_i}\}$ is chosen here to be direct.

The electric field circulating inside the cavity and the cavity enhancement factor $G$ are given by:

$$\mathbf{E}_{circ} \propto \left[\sum_{n=0}^{\infty} \left(Je^{i\psi}\right)^n\right] t_1 \mathbf{V_0}, \quad G = \left\|\left[\sum_{n=0}^{\infty} \left(Je^{i\psi}\right)^n\right] t_1 \mathbf{V_0}\right\|^2 \tag{0.4}$$

where $t_1$ is the 2×2 transmission matrix of the injection mirror 1, $\psi = 2\pi\Lambda/\lambda$ is the part of round-trip phase shift which is independent of $\varphi_{ip}$ and $\varphi_{is}$ [1] and $\mathbf{V_0}$ is the polarization vector of the incident laser beam ($\mathbf{V_0} = (1,i)/\sqrt{2}$ and $\mathbf{V_0} = (1,0)$ for a circularly and linearly polarized beam respectively). The series is conveniently calculated in the eigenvector basis $\{\mathbf{e_1}, \mathbf{e_2}\}$ which diagonalizes $J$. One obtains

$$\mathbf{E}_{circ} \propto U \begin{pmatrix} \dfrac{1}{1-\xi_1 e^{i\zeta_1} e^{i\psi}} & 0 \\ 0 & \dfrac{1}{1-\xi_2 e^{i\zeta_2} e^{i\psi}} \end{pmatrix} U^{-1} \cdot t_1 \cdot \mathbf{V_0} \tag{0.5}$$

In eq. (0.5), U is the matrix changing to the basis $\{\mathbf{e_1}, \mathbf{e_2}\}$, that is

$$U = \begin{pmatrix} \mathbf{s_1} \cdot \mathbf{e_1} & \mathbf{s_1} \cdot \mathbf{e_2} \\ \mathbf{p_1'} \cdot \mathbf{e_1} & \mathbf{p_1'} \cdot \mathbf{e_2} \end{pmatrix} \tag{0.6}$$

and $\xi_1 \exp(i\zeta_1)$ and $\xi_2 \exp(i\zeta_2)$ are the two eigenvalues of $J$, i.e.

$$U^{-1}JU = \begin{pmatrix} \xi_1 \exp(i\zeta_1) & 0 \\ 0 & \xi_2 \exp(i\zeta_2) \end{pmatrix} \tag{0.7}$$

When the cavity mirrors are misaligned, one has $\zeta_1 \neq \zeta_2$ which means that the two eigenvectors exhibit different resonance frequencies. Since a cavity is locked on a unique frequency, one is free to choose between the two eigenmodes. To perform our numerical choice we have considered the following simple algorithm which can be put into practice:

$$\text{If } |\mathbf{e_1} \cdot t_1 \mathbf{V_0}| > |\mathbf{e_2} \cdot t_1 \mathbf{V_0}| : \mathbf{E_{circ}} \propto U \begin{pmatrix} \dfrac{1}{1-\xi_1} & 0 \\ 0 & \dfrac{1}{1-\xi_2 e^{i(\zeta_2-\zeta_1)}} \end{pmatrix} U^{-1} t_1 \mathbf{V_0}$$

$$\text{If } |\mathbf{e_2} \cdot t_1 \mathbf{V_0}| \geq |\mathbf{e_1} \cdot t_1 \mathbf{V_0}| : \mathbf{E_{circ}} \propto U \begin{pmatrix} \dfrac{1}{1-\xi_1 e^{-i(\zeta_2-\zeta_1)}} & 0 \\ 0 & \dfrac{1}{1-\xi_2} \end{pmatrix} U^{-1} t_1 \mathbf{V_0}$$

Finally, the Stokes vector components are computed from the expression of the circulating beam:

$$\begin{aligned} S_1 &= \dfrac{|\mathbf{E_{circ}} \cdot \mathbf{s}|^2 - |\mathbf{E_{circ}} \cdot \mathbf{p}|^2}{\|\mathbf{E_{circ}}\|^2} \\ S_2 &= \dfrac{(\mathbf{E_{circ}} \cdot \mathbf{s})(\mathbf{E_{circ}} \cdot \mathbf{p})^* + (\mathbf{E_{circ}} \cdot \mathbf{s})^* (\mathbf{E_{circ}} \cdot \mathbf{p})}{\|\mathbf{E_{circ}}\|^2} \\ S_3 &= \dfrac{i\left[(\mathbf{E_{circ}} \cdot \mathbf{s})(\mathbf{E_{circ}} \cdot \mathbf{p})^* - (\mathbf{E_{circ}} \cdot \mathbf{s})^* (\mathbf{E_{circ}} \cdot \mathbf{p})\right]}{\|\mathbf{E_{circ}}\|^2} \end{aligned} \tag{0.8}$$

## Section 3 Numerical results

The numerical computations have been performed for cavity designs fulfilling the requirements of the laser Compton experiments described in section 1. More specifically, we consider a cavity with $L$=500mm, $h$=100mm and a laser beam of wavelength $\lambda$=1030nm. For the Z folded planar and non planar cavities we further take $d$=250mm. These numbers correspond to a round-trip length of ~2m. The radii of curvatures of mirrors 3 and 4 are taken to be the same. We further impose $R=\|\overrightarrow{M_{c3}M_{c4}}\|\cos(\theta_0)$ which corresponds to the smallest cavity waist [27].

*Mechanical tolerances*

In order to estimate the numerical tolerances, the optical axis is computed for all the $2^{20}$ combinations of misalignments parameters $\Delta r_i = \{-1,+1\}$ µrad, with $\Delta r_i = \Delta X_i, \Delta Y_i, \Delta Z_i$ and $\Delta \Theta_i = \{-1,+1\}$ µm, with $\Delta \Theta_i = \Delta \theta_{xi}, \Delta \theta_{yi}$ ($i$=1,..,4). These values are arbitrary and will be related to the tolerance parameter defined below.

For each configuration, we record the distances $\|\overrightarrow{M_{ci}M_i}\|$ between the mirror centers $M_{ci}$ of mirror $i$ and the reflection point $M_i$ of the optical axis on this mirror. The maximum distance $\|\overrightarrow{M_{ci}M_i}\|$ among the $2^{20}$ configurations is denoted by $\Delta_{\max}$ and is considered as the tolerance length parameter.

We obtain a tolerance length 9µm for the bow-tie planar and tetrahedron non planar cavities. Keeping $L$ unchanged and varying $h$ from 100mm to 200mm does not change significantly the tolerances (one gets 9.5µm instead of 9µm) and changing $L$ from 500mm to 1000mm increases the tolerance to 12µm.

It is interesting to discuss here in greater detail the case of the U and Z folded planar cavities, in view of the results of ref. [28] where it has been shown that the condition $R=\|\overrightarrow{M_{c3}M_{c4}}\|\cos(\theta_0)$ corresponds to an instable configuration. Contrary to ref. [28], where Z-folded cavities were considered for dye lasers, in our setting we are free to modify the distance $D$ between the flat and curve mirrors $D=\left((h/2)^2+d^2\right)^{1/2}$ (see Figure 3). Figure 5 shows the tolerance length $\Delta_{max}$ as a function of $D$ for the four planar and non planar, U and Z folded geometries. To draw these curves, we only considered tilt misalignments $\Delta\Theta_i=\{-1,+1\}$ µm (with $\Delta r_i=0$) and we fixed $L$=1000mm and $\theta_0=\pi/4$ for the Z-folded planar and non planar configurations. One sees from the figure that for large D, non planar configurations are much more mechanically stable than the planar ones. One also sees that U folded cavities are more stable than Z-folded ones, a better mechanical stability being reached for non planar U folded cavities. We have checked that, in all cases, $\Delta_{max}$ decreases when $R<\|\overrightarrow{M_{c3}M_{c4}}\|\cos(\theta_0)$ as observed in [28].

From the point of view of mechanical stability, it comes out that all of the U and Z folded geometrical configurations can be considered provided the cavity parameters $h$, $d$ and $\theta_0$ are carefully chosen. Among these configurations, U folded non planar cavities offer interesting geometrical features when the implementation of an optical four-mirror cavity on an electron accelerator is envisaged. However, one drawback is that the corresponding cavity eigenmodes are strongly elliptical.

*Polarization and enhancement factor stability*

We shall now numerically estimate the sensitivity of the polarization eigenvectors and of the circulating field to cavity mirror misalignment and motions. The cavity length parameters $L$ and $d$, the laser beam wavelength and the mirror radius of curvature $R$ are set as the same values as in the previous section. Three values of the parameters $h$=50,100,200 mm are considered here. We adopt the following numerical procedure:

> ➤ First, the optical axis is computed as in the previous section for a given set of tilt misalignment angles $\Delta\Theta_i$ for $i$=1,..,4 and we further set $\Delta r_i = 0$ in order to save computer time.

> ➤ The Jones matrix is then computed leading to the eigenvectors and enhancement factor of the cavity.

These two steps are first performed for each combination of the tilt misalignments angles $\Delta\Theta_i = \{-500, 0, +500\}$ μrad with $i$=1,..,4, *i.e.* $3^8$ configurations. Note that the choice of 500μrad for the misalignment angle corresponds to the typical residual misalignment of a long Fabry-Perot cavity. Since we are interested in applications where the laser beam inside the cavity is circularly polarized, we assume that the polarization vector of the incident laser beam is $\mathbf{V_0} = (1, i)/\sqrt{2}$. As for the number of double layers constituting the mirror coating, we choose $N$=4, 12, 20 for mirrors 2, 3, 4 and $N$-2 for the entrance mirror 1 in order to account for the cavity phase matching.

After computing the degree of circular polarization $S_3$ of eq. (0.8) and the cavity enhancement factor $G$ of eq. (0.4) for each combination of tilt misalignments, we obtain the two ensembles $\{S_3\}$ and $\{G\}$ from which we calculate the following simple statistical estimators: the averages $<S_3>$=mean($\{S_3\}$), $<G>$=mean($\{G\}$), the root mean squares $\sigma(S_3)$=rms($\{S_3\}$), $\sigma(G)$=rms($\{G\}$) and the maximum spread $\Delta(S_3)$=Max($\{S_3\}$)-Min($\{S_3\}$), $\Delta(G)$=Max($\{G\}$)-Min($\{G\}$). We have numerically checked that for values of $L$ up to 2m (with $R=\|\overrightarrow{M_{c3}M_{c4}}\|\cos(\theta_0)$) the previous estimators only depend on the ratio $h/L$ as expected since the effects studied in this article are related to angles of incidence on the cavity mirrors (in fact these effect also depends on the mechanical tolerance, but at a negligible level in our numerical cases). We shall therefore show our numerical results as a function of the single parameter

$$e = \frac{h}{2L}$$

instead of $L$ and $h$ separately.

We start to show our numerical results by comparing the stability of the 2D bow-tie and 3D tetrahedron cavities. Figure 6, Figure 7, Figure 8 and Figure 9 show $<G>$, $<S_3>$, $\Delta(G)/<G>$ and $\Delta(S_3)$ as a function of the number of coating double layers $N$ for $e$=0.4, 0.2, 0.1. From these figures, one sees that the averages $<G>$ and $<S_3>$ are not strongly affected by the misalignments, whereas the values of $\Delta(S_3)$ and $\Delta(G)$ can be very large. In fact, Figure 8 and Figure 9 illustrate that $\Delta(G)$ and $\Delta(S_3)$ are negligible for all the values of $e$ and $N$ in the 3D case. However, for the 2D case, these figures show that $\Delta(S_3)$ and $\Delta(G)$ are large for large values of $N$ or $e$ and negligible for all $N$ when $e$ 0.1.

We have chosen to discuss $\Delta(G)$ and $\Delta(S_3)$ instead of $\sigma(G)$ and $\sigma(S_3)$ because they are the relevant quantities for the applications described in section 1. The latter estimators are indeed smaller by a factor of ~ 5 as can be seen by comparing Figure 8 to Figure 10 where the values of $\sigma(G)/<G>$ are plotted.

To pin down the origin of the instability, *i.e.* the large values of Δ($S_3$) and Δ($G$), the eigenvector polarizations are shown in Figure 11 for all the misalignment configurations of a 2D cavity with $e$=0.2 and $N$=4, 12, 20. The Poincaré sphere representation of these polarization vectors [29] is adopted here with the following choice for the polar angle $\theta_{Ps} = \cos^{-1}(S_{3,e})$ and azimuth angle $\phi_{Ps} = \tan^{-1}(S_{2e}/S_{1e})$ ($S_{ie}$ are the components of the stokes vector corresponding to the polarization vectors **e₁** or **e₂** of the cavity eigenmodes). From Figure 11 one thus sees that the eigenmodes are linearely polarized (i.e. $\theta_{Ps}$=π/2 and $\phi_{Ps}$=0,π which correspond to $S_1$=±1) for $N$=4 and that they become more and more elliptical as $N$ increases. The instabilities observed in Figure 8 and Figure 9 for the 2D cavities are thus related to the variations of the eigenvector polarization with the misalignment angles. As for the 3D cavities, it comes out that the eigenvectors are always circularly polarized (*i.e.* $\theta_{Ps}$=0,π for all the misalignment configurations). The instabilities of $G$ and $S_3$ are therefore induced by the coupling of the incident polarization vector **V₀** with the cavity eigenvectors **e₁** and **e₂** which represent the polarisation of the cavity eigenmodes.

We turn now to the study of mirror motions. The same study for $\Delta\Theta_i = \{-5, 0, +5\}$ μrad with $i$=1,..,4 is performed (note that for a one inch diameter mirror mounted in a gimbal mount, a tilt of 5 μrad corresponds to a vibration amplitude of ~75nm of the mirror edge with respect to its centre). The corresponding values of Δ($G$) and Δ($S_3$) are shown in Figure 12 and Figure 13 respectively. By comparing these figures with Figure 8 and Figure 9, one sees that the instability reduction is not enough for large value of $N$ where a simple scaling by a factor of 1/100 does not hold. This figure shows that if one wants, to reduce the polarization instabilities induced by the mirror motion of a very high finesse cavity below the percent level, a 3D tetrahedron geometry or a 2D geometry with $e<0.1$ must be used.

We also numerically investigated the effect of the incident polarization and of the laser beam wavelength. Considering an incident linearly polarized laser beam, *i.e.* $\mathbf{V_0} = (1,0)$, we observed qualitatively the same instabilities for the 2D geometry and still a high degree of stability for the 3D geometry. Changing the wavelength from 1030 nm to 515 nm, we found that the instabilities increase by a factor of ~5. The polarization instabilities induced by the cavity mirror misalignments or motion therefore occur whatever the polarization and wavelength of the incident radiation, but at a different degree.

Finally, we also performed the same numerical study for U and Z folded cavities. We set the geometrical parameters $d$=250 mm and $L$=500 mm and varied the number of coating double layers $N$ and the length parameter $h$ as described above. As a result, we obtained results similar to the planar bow tie cavity ones. Here the non planar extensions of the Z and U folded cavities do not reduce the instabilities. The reason is that the optical axis is always self-retracing so that the eigenmodes are linearly polarized for low finesse and become slightly elliptical as the finesse increases as in the 2D geometry case.

### *Eige modes of the tetrahedron cavity*

While the shape of a planar four-mirror cavity is well known [27], the one of a tetrahedron has not been described yet. Therefore, we now investigate the shape of the fundamental eigenmode of the tetrahedron cavity. The eigenmode of such a cavity belongs, in the paraxial approximation, to the class of general astigmatic beams [31] (its intensity profile is elliptical and the orientation of the ellipse axes is changes during the beam propagation). Such modes are indeed numerically computable using the formalism of [32]. To illustrate the properties of the fundamental mode, we use the following numerical values: $h$=100 mm, $R$=500 mm, $\lambda$=1 µm and $L$=495.2 mm and 511mm. The corresponding beam radii $\omega_1$ and $\omega_2$ along the major and minor ellipse axes and $\alpha$, the ellipse orientation angle in the $\{s_3,p_3\}$ basis, are shown in Figure 14 and Figure 15 respectively as a function of the unfolded coordinate along the mode propagation axis $z_{beam}$. We respectively obtain the beam waists $\omega_{01}$=32 µm ($\omega_{01}$=97.3 µm) and $\omega_{02}$=53 µm ($\omega_{02}$=97.5 µm) between the two spherical mirrors for $L$=495.2 mm ($L$=511 mm). As expected [31], Figure 15 shows a fast rotation of the ellipse close to the waist position. In addition, as also expected [32], a full $\pi$ rotation of the angle $\alpha$ is obtained during a cavity round-trip and one further sees that for small waits this rotation occurs almost completely between the two spherical mirrors. Figure 14, shows that the beam ellipticity strongly decreases as the waists increase so that for $L$=511 mm the intensity profile is almost circular.

The main difference between the bow-tie and tetrahedron cavity is the rotation of the elliptical intensity profile which is only noticeable when the waists are small. In this case, as we are interested colliding the laser beam onto an electron beam of typical longitudinal length ~1mm, Figure 14 shows that the ellipse rotation is small within such a distance so that a good enough overlap between the beams can be kept. A quantitative estimate of this effect on the laser-electron beams luminosity is outside the scope of the article and will be reported on elsewhere.

## Section 4 Summary

We have investigated the stability of various geometrical configurations of four-mirror cavities in the context of future X and gamma rays Compton machines. We indicated that stringent constraints are indeed put on the geometrical design and operation stability for the applications envisaged for these machines.

We have numerically shown that the polarization coupling of the incident laser beam with the four-mirror cavity eigenmodes induces an enhancement factor and polarization instabilities when mirror misalignments motions are taken into account. We observed that this instability depends on the ratio of the 'cavity width' to the 'cavity length' *e=h/2L* (see Figure 1) and not on *h* and *L* independently.

For planar bow tie and Z folded geometries, these instabilities are small when the angles of incidence on the mirrors and the cavity finesse are kept small enough, that is when *e<0.1*. They increase non linearly when the cavity finesse increases and, for a given finesse, they decrease when the angle of incidence decreases. The latter feature leads to an incompatibility with the mechanical stability conditions of U and Z folded cavities which worsen when the incidence angle decreases. The design of high finesse and highly stable U or Z folded planar resonators may therefore prove difficult, whereas stable bow-tie cavities can be considered provided that the condition *e<0.1* is fulfill .

We have studied non planar extensions of the bow tie and Z folded planar cavities. We found that while the non planar Z folded geometry does not reduce the polarization instabilities, the tetrahedron geometry does reduce them at a very small level for all the values of the parameter *e*. This configuration must then be experimentally studied to provide a good technical solution for the applications described in the introduction. This is a Research and Development activity which has started in our Laboratory and that we shall report on in the near future.

One aspect of the cavity coating which has not been tackled in this article is their residual birefringence [30]. Although this very small effect should have noticeable effects for very high finesse cavities, we did not find a robust method to include them in our numerical studies.

## References

1. H. Kogelnick and T. Li, "Laser beams and resonators", *Appl. Opt.*, vol. 5, pp. 1550-1567, 1966.

2. RJ. Jones, JC. Diels, J. Jasapara and W. Rudolph, "Stabilisation of the frequency, phase, repetition rate of an ultra-short pulse train to a Fabry-Perot cavity", *Opt. Comm.,* vol. 175, pp. 409-418, 2000.

3. P. Sprangle, A. Ting, E. Esarey, and A. Fisher, "Tunable, short pulse hard X-rays from compact laser synchrotron source", J. Appl., Phys., vol. 72, pp. 5032-5034, 1992.

4. J. Chen, K. Iinasaki, M. Fujita, C. Yamanaka, M. Asakawa, S. Nakai and T. Asakuma, "Development of a compact high brightness X-ray source", Nucl. Instr. Meth. A, vol. 341, pp. 346-350, 1994.

5. Z. Huang and R.D Ruth, "Laser-electron storage ring"Phys, Rev. Lett., vol. 80, pp. 976-979, 1998.

6. F.E. Carroll, "Tunable Monochromatic X Rays: A New Paradigm in Medicine", American Journal of Roentgenology, vol. 179, pp. 583-590, 2002.

7. M.-C. Biston, A. Joubert, J.-F. Adam, H. Elleaume, S. Bohic, A.-M. Charvet, F. Estève, N. Foray and J. Balosso, "Cure of Fisher Rats Bearing Radioresistant F98 Glioma Treated with cis-Platinum and Irradiated with Monochromatic Synchrotron X-Rays", Cancer Research, vol. 64, pp. 2317-2323, 2004.

8. P. Suorti and W. Thomlinson, "Medical applications of synchrotron radiation", Phys. Med. Biol., vol. 48, R1-R35, 2003.

9. M. Cotte, E. Welcomme, V. A. Solé, M. Salomé, M. Menu, Ph. Walter and J. Susini, "Synchrotron-Based X-ray Spectromicroscopy Used for the Study of an Atypical Micrometric Pigment in 16th Century Paintings", Anal. Chem., vol. 79, pp. 6988-6994, 2007.

10. R. Hajima, T. Hayakawa, N. Kikuzawa and E. Minehara, "Proposal of Nondestructive Radionuclide Assay Using a High-Flux Gamma-Ray Source and Nuclear Resonance Fluorescence", J. Nucl. Sci. Tech., vol. 45, pp. 441-451, 2008.

11. G. Moortgat-Pick et al., "The Role of polarized positrons and electrons in revealing fundamental interactions at the linear collider", Physics Reports, vol. 460, pp. 131–243, 2008.

12. R.J. Loewen, "A compact light source: design and technical feasibility study of a laser-electron storage ring X-ray source," Ph.D. dissertation, Dept. Phys., Stanford Univ., Stanford, CA, 2003.

13. C.J. Hood, H.J. Kimble and J. Ye, "Characterization of high-finesse mirrors: loss, phase shifts, and mode structure in optical cavity", Phys. Rev. A, vol. 64, pp. 033804-033811, 2001.

14. N.I. Zheludev, "Polarization instability and multistability in nonlinear optics", Sov. Phys. Usp., vol. 32, pp. 357-375, 1989.

15. D.T. Atwood, *Soft X-rays and extreme ultraviolet radiation*, Cambridge University Press, Cambridge 2007.

16. S. Balestri, P. Burlamacchi, V. Greco and G. Molesini, "Folded $CO_2$ laser resonators with controlled beam quality", Opt. Comm., vol. 104, pp. 91-06, 1993.

17. A.E. Siegman, *Lasers*, CA: University Science Books, Sausalito 1986, p. 607.

18. J. Yuan and X. Long, "Optical-axis perturbation in nonplanar ring resonators", Opt. Comm., vol. 281, pp. 1204-1210, 2008.

19. S.A. Collins, Jr., "Analysis of optical resonators involving focusing elements", Appl. Opt., vol. 64, pp. 1263-1275, 1964.

20. Matlab 6.5 software, The MathWorks Inc., 3 Apple Hill Drive, Natick, United States of America.

21. J.A. Arnaud, "Degenerate optical cavity", Appl. Opt., vol. 8, pp. 189-195, 1969.

22. A.C. Nilsson, E.K. Gustafson and R.L. Byer, "Eigenpolarization theory of monolithic nonplanar ring oscillators", IEEE J. Quant. El., Vol. 25, pp. 767- 790, 1989.

23. E. Hecht, *Optics*, Addison Wesley, San-Francisco 2002, p 426.

24. S. Saraf, R.L. Byer and P.J. King, "High-extinction-ratio resonant cavity polarizer for quantum-optics measurements", Appl. Opt., vol. 46, pp. 3850-3855, 2007.

25. W.W. Chow, J. Gea-Banacloche, L.M. Pedrotti, V.E. Sanders, W. Schleich and M.O. Scully, "The ring laser gyro", Rev. Mod. Phys., vol. 57, pp. 61-104, 1985.

26. H. Jiao, S.R. Wilkinson, R.Y. Chiao and H. Nathel, "Topological phases in optics by means of nonplanar Mach-Zehnder interferometer", Phys. Rev. A, Vol. 39, pp. 3475-3486, 1989.

27. H.W Kogelnik, E.P. Ippen, A. Dienes and C.V. Shank, "Astigmatically compensated cavities for cw dye lasers", IEEE J. Quant. El., vol. QE-8, pp. 373-379, 1972.

28. V. Magni, S. De Silvestri and A. Cybo-Ottone, "On the stability, mode properties, and misalignment sensitivity of femtosecond dye laser resonators", Opt. Comm., vol. 82, pp. 137-144, 1991.

29. M. Born and E. Wolf, *Principle of Optics*, Pergamon Press, Oxford 1965, p 30.

30. D. Jacob, M. Vallet, F. Bretenaker, A. Le Floc and M. Oger, "Supermirror phase anisotropy measurement", Opt. Lett., vol. 20, pp. 671-673, 1995.

31. J.A. Arnaud and H. Kogelnik, "Gaussian light beam with general astigmatism", Appl. Opt., vol. 8, pp. 1687-1693, 1969

32. J.A. Arnaud, "Nonorthogonal waveguides and resonators", *Bell Syst. Tech. J.*, pp. 2311-2348, 1970.

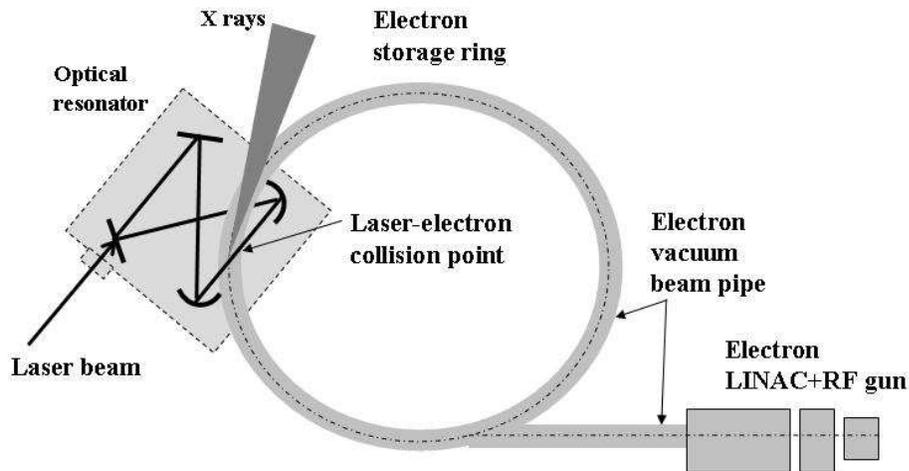

**Figure 1 Schematic design of an electron ring accelerator and a four-mirror optical cavity used to produce X rays by Compton scattering.**

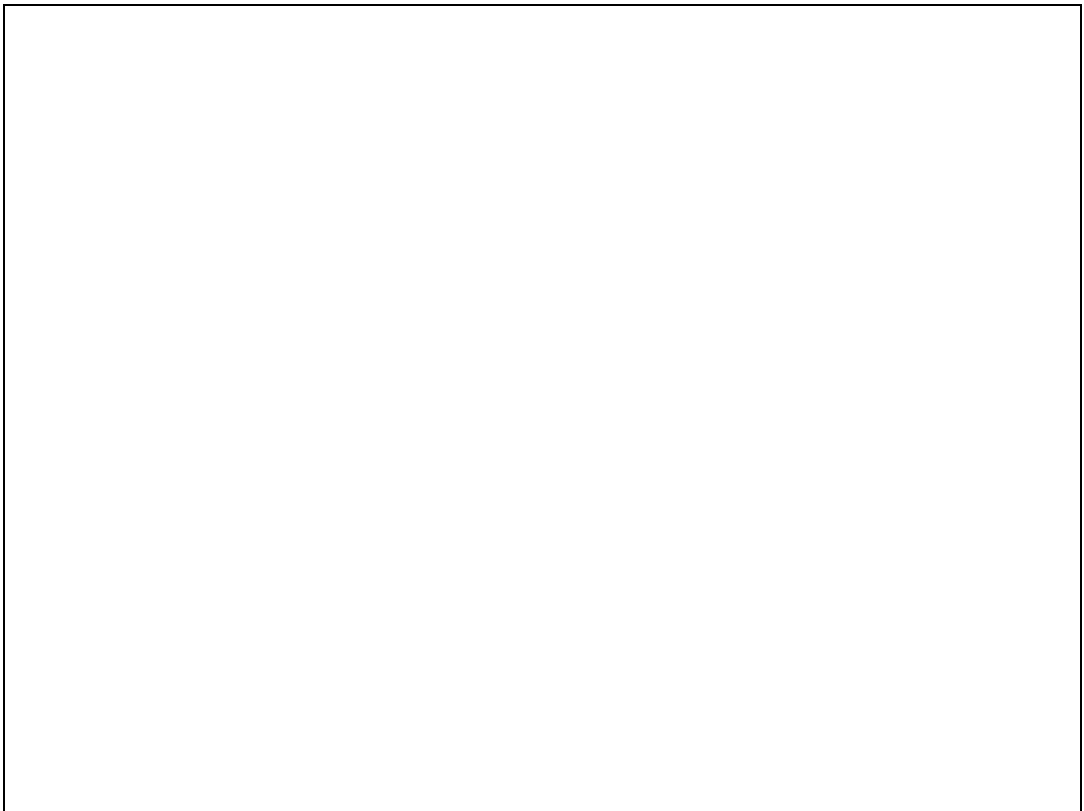

**Figure 2 (a) the normalised angular emission of Compton scattering is displayed for the cases γ = 10,100,300. (b) points out the laser wavelength dependence of the emitted photons energy cut for an electron beam of 50 MeV and for laser photon energies of 1 eV and 2 eV.**

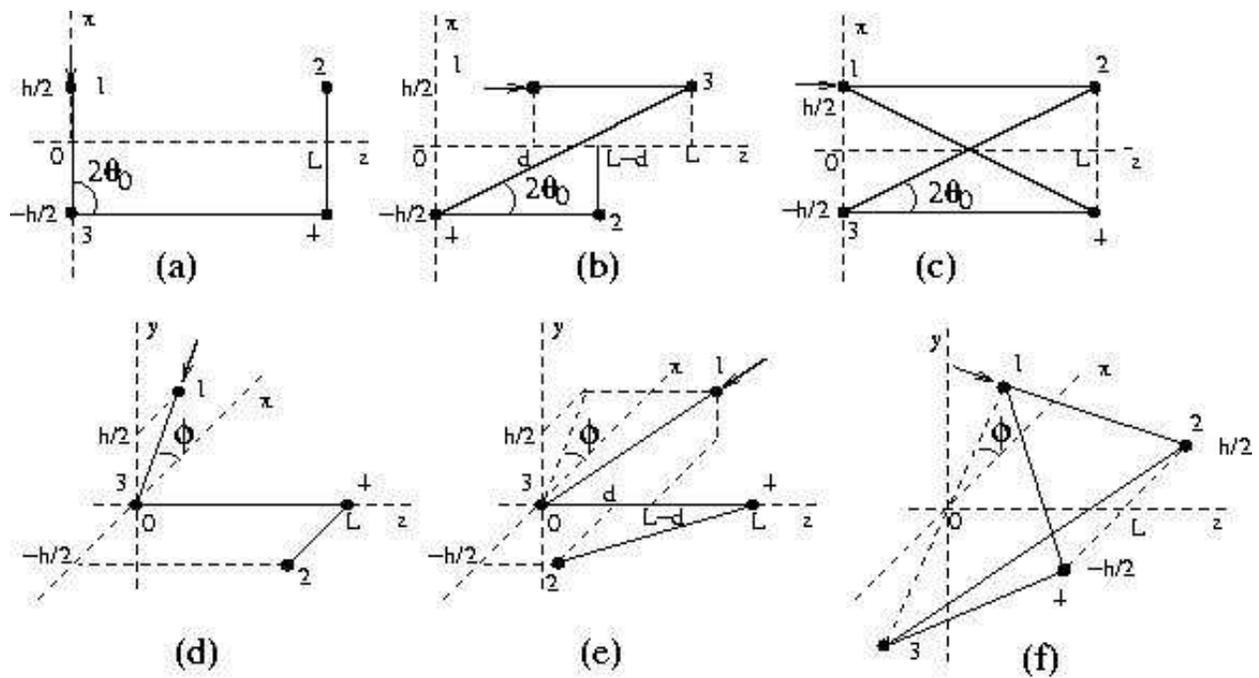

**Figure 3** Cavity geometries: a) planar U folded, b) planar Z folded, c) planar bow-tie, d) non planar U folded type, e) non planar Z folded type, f) non-planar bow-tie type. Numbers 1 and 2 indicate the locations of the flat mirrors and numbers 3 and 4 the locations of spherical mirrors. The reference axes $x$, $y$, $z$ are shown.

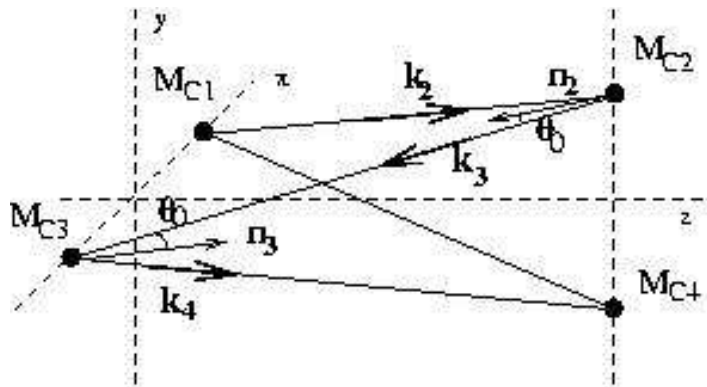

**Figure 4** Tetrahedron non planar cavity together with the wave vectors and normal vectors of mirrors 2 and 3. The points $M_{Ci}$ correspond to mirror centers.

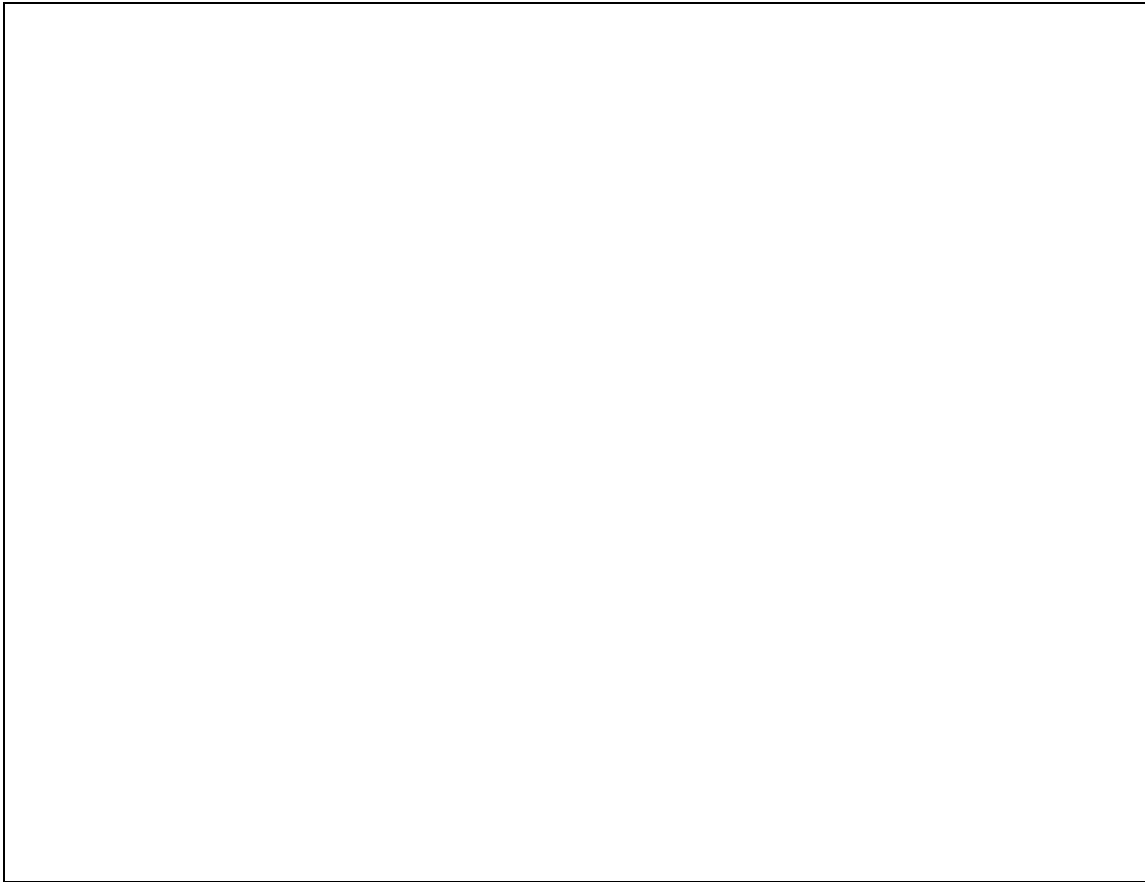

**Figure 5 Tolerance lengths as a function of the distance between the flat and spherical mirrors for the U and Z folded, planar and non planar geometries. Only mirror tilting misalignments are taken into account to compute Δ$_{max}$ (see text).**

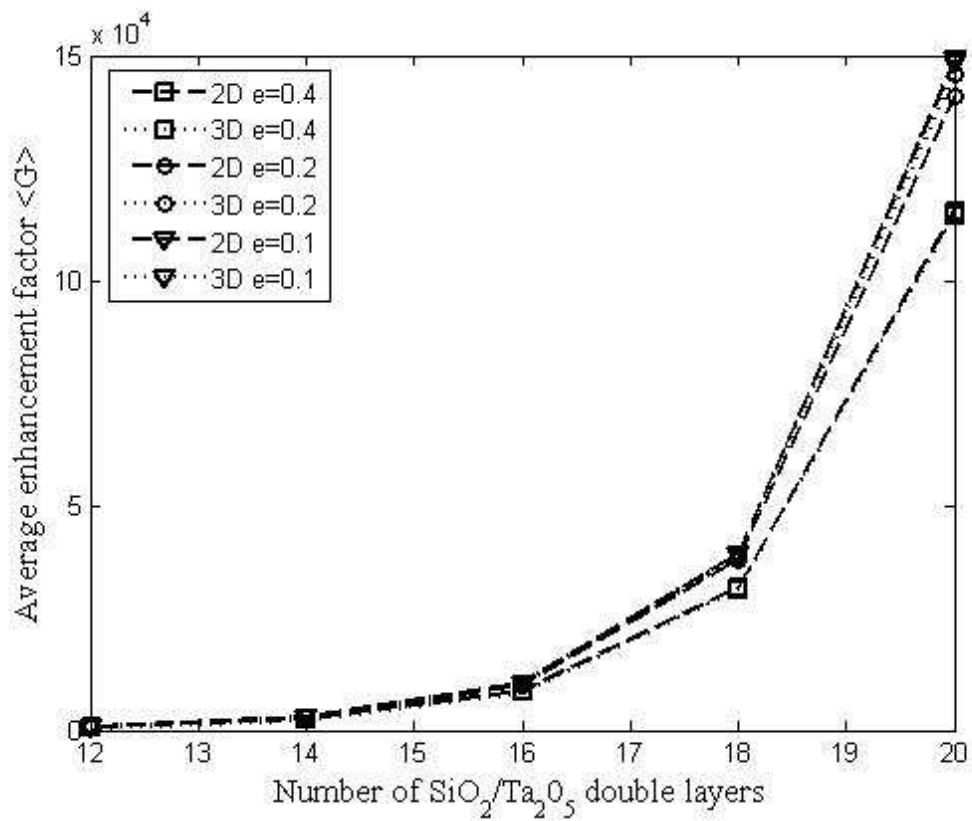

**Figure 6 Average enhancement factor *<G>* over $2^8$ misalignment configurations (see text) as a function of the number of double layers of the mirror coatings.**

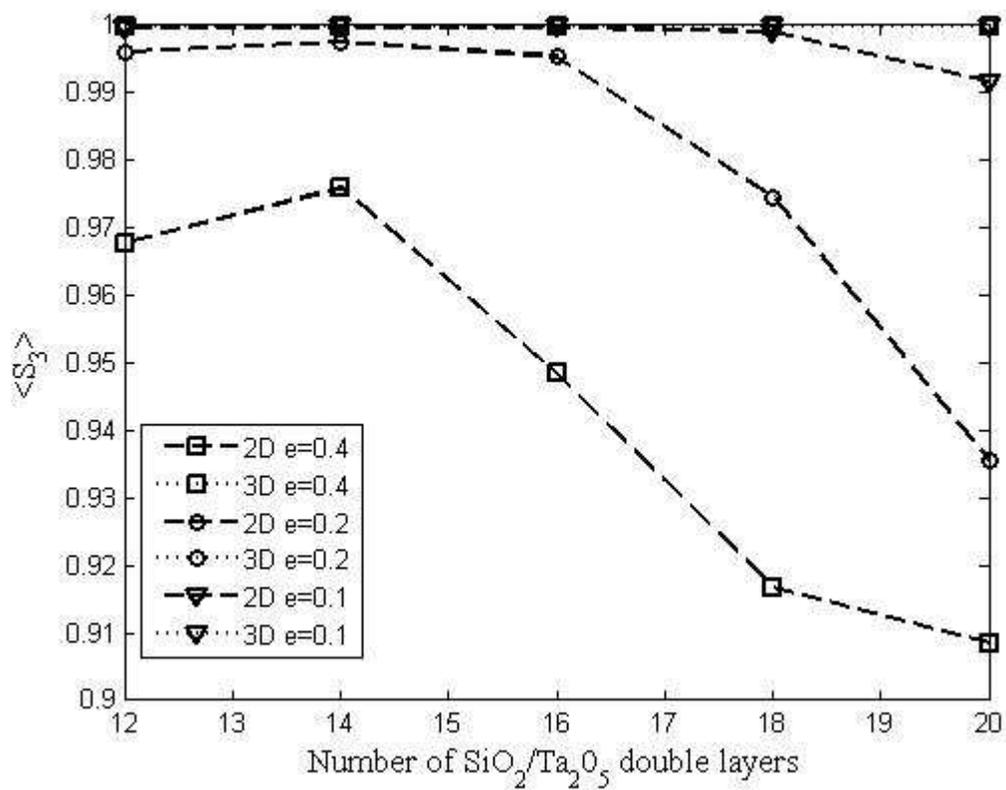

**Figure 7** Average degree of circular polarization $<S_3>$ over $2^8$ misalignment configurations (see text) as a function of the number of double layers of the mirror coatings.

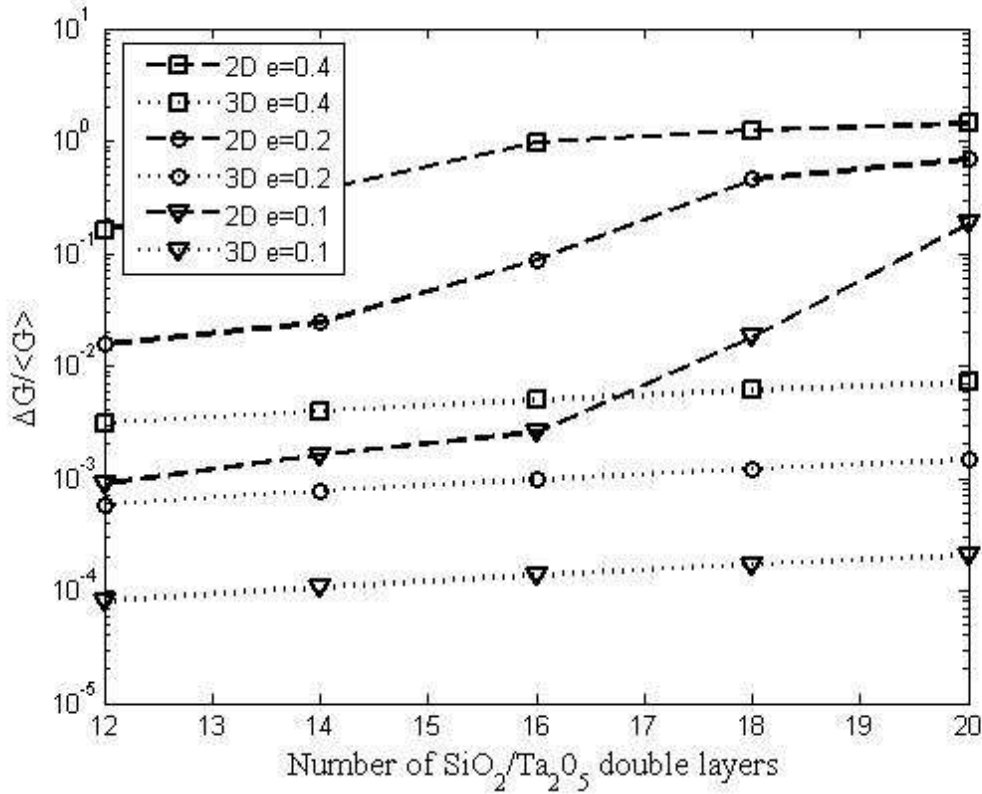

**Figure 8** $\Delta G/\langle G \rangle$ **corresponding to the $2^8$ misalignment configurations (see text) as a function of the number of double layers of the mirror coatings.**

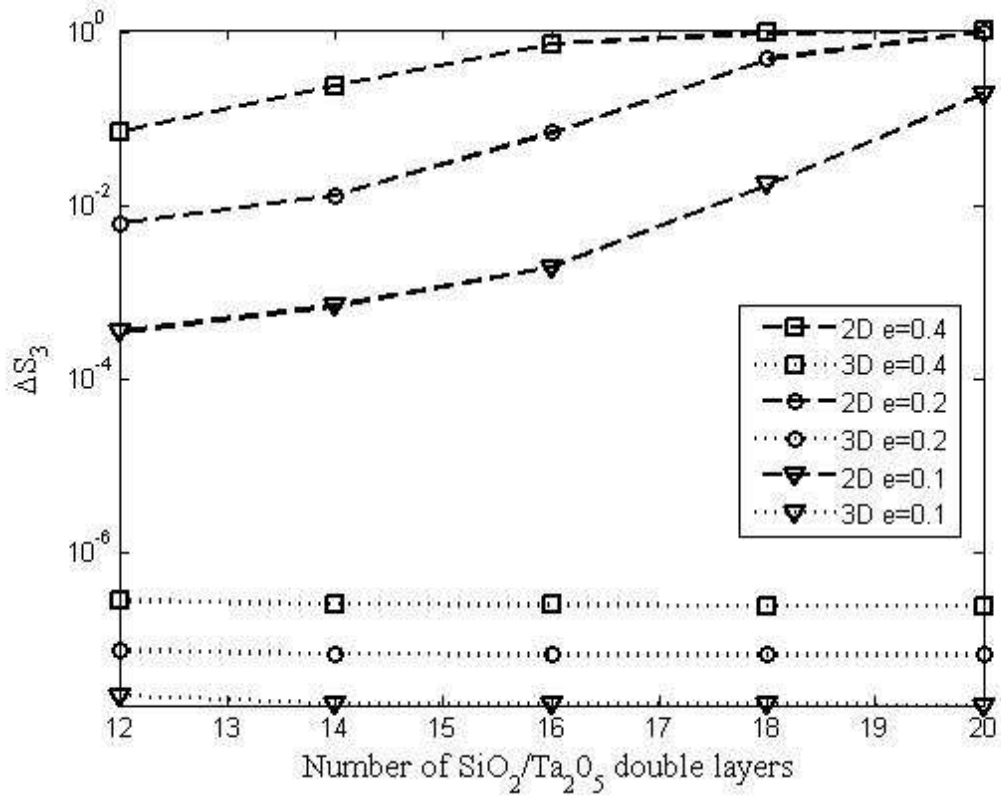

**Figure 9** $\Delta S3$ **corresponding to the $2^8$ misalignment configurations (see text) as a function of the number of double layers of the mirror coatings.**

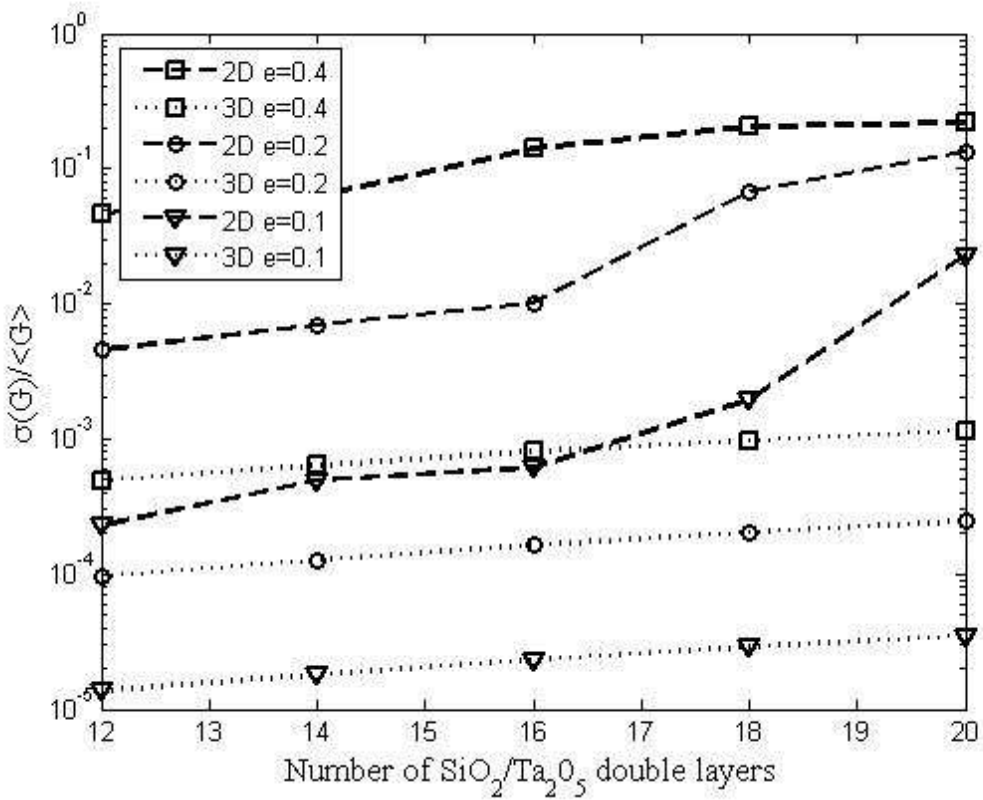

**Figure 10** $\sigma(G)/<G>$ > corresponding to the $2^8$ misalignment configurations (see text) as a function of the number of double layers of the mirror coatings.

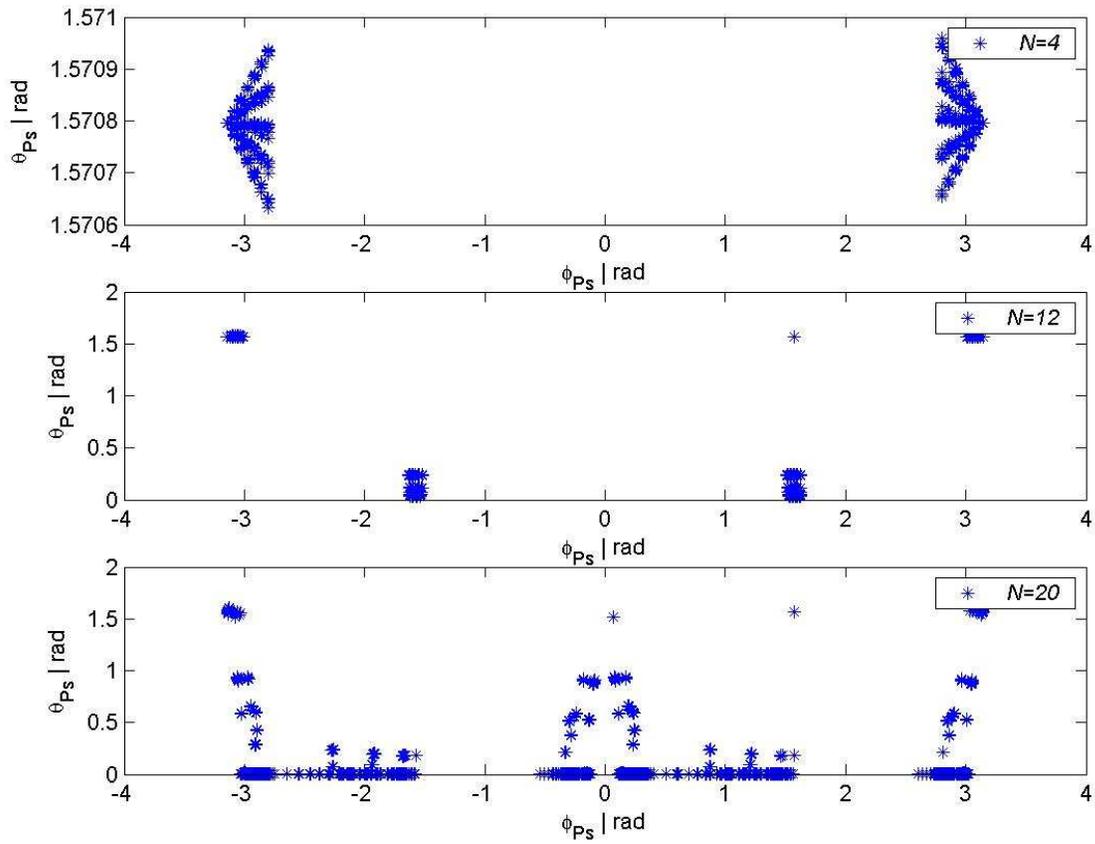

**Figure 11 Eigenvector representations on the Poincaré sphere for a Bow-tie cavity and various number of mirror coating double layers *N*. The points correspond to the $2^8$ misalignment configurations (see text). The geometrical parameters are fixed to *L*=500mm and *h*=100mm.**

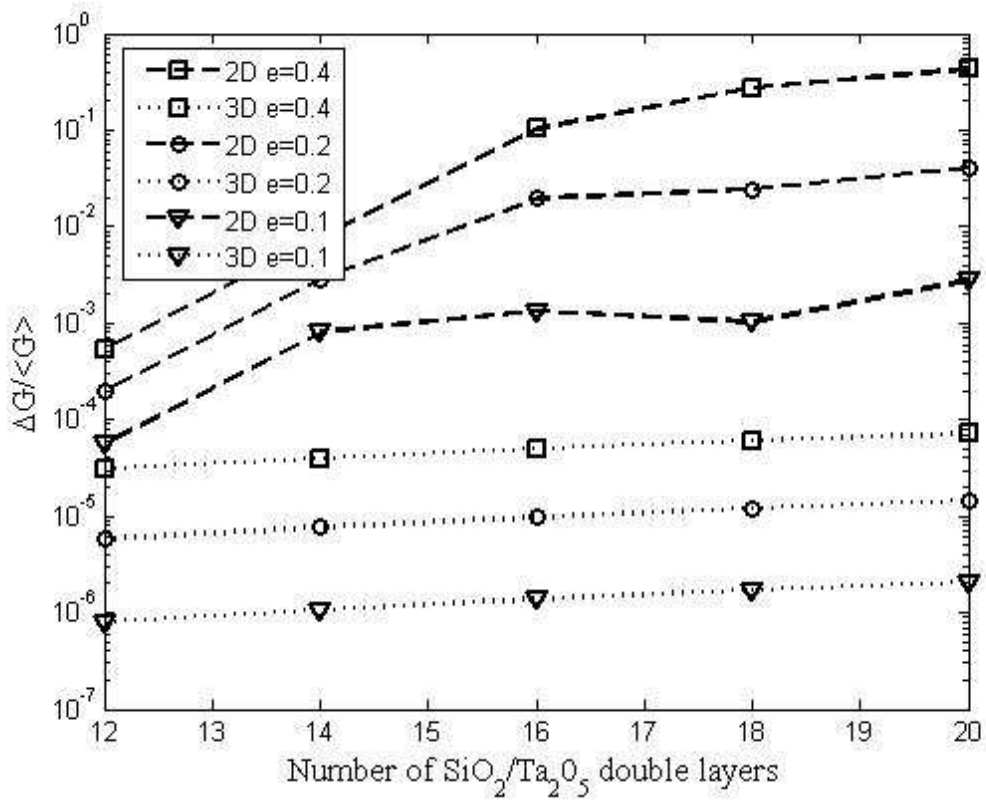

**Figure 12** $\Delta G/<G>$ **corresponding to the $2^8$ mirror motion configurations (see text) as a function of the number of double layers of the mirror coatings.**

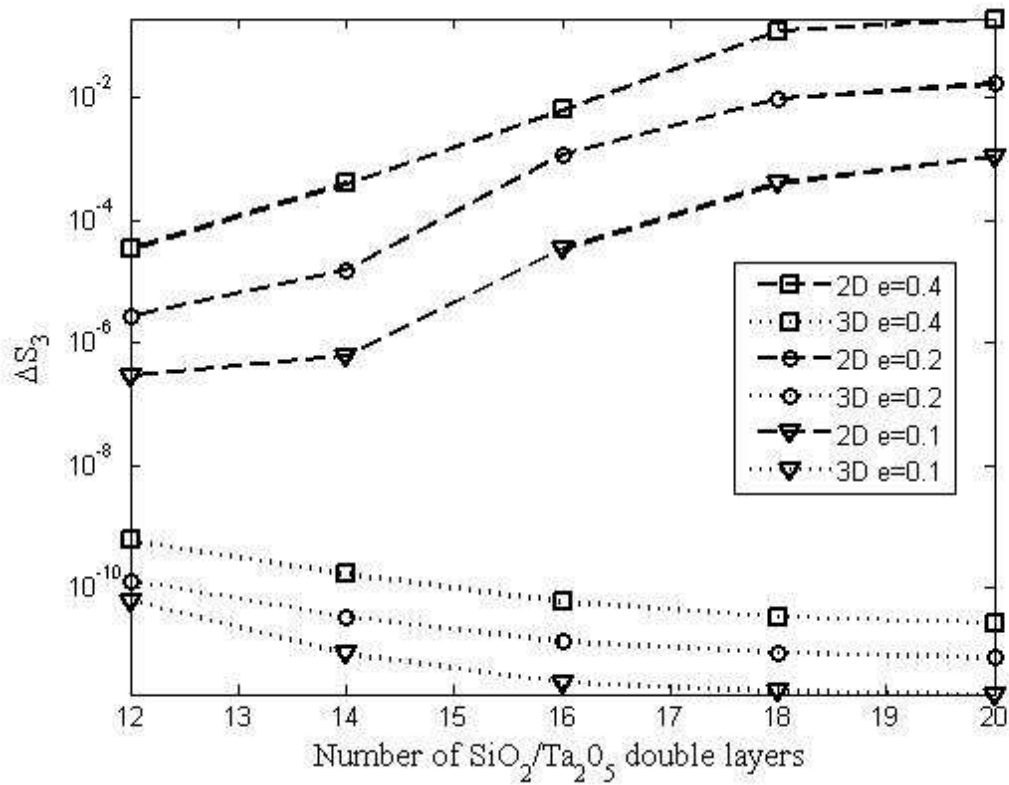

**Figure 13 $\Delta S_3$ corresponding to the $2^8$ mirror motion configurations (see text) as a function of the number of double layers of the mirror coatings.**

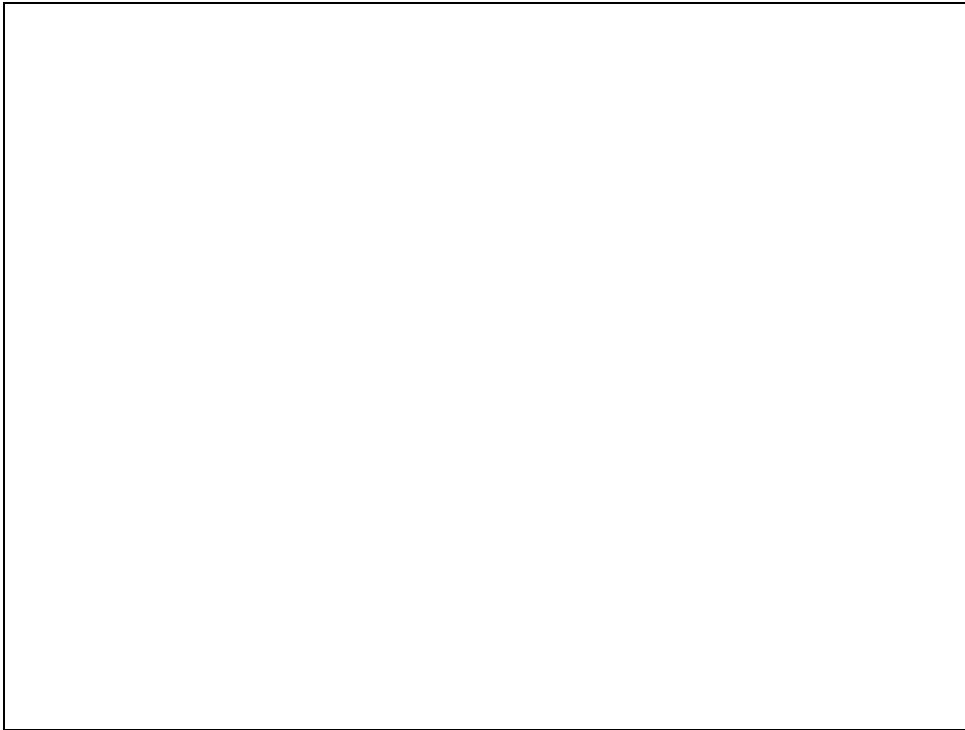

**Figure 14 Beam radii of the fundamental mode of two tetrahedron cavities (see text) as a function of the unfolded coordinate along the mode propagation axis $z_{beam}$. The dotted and the full lines are located on each other. The positions $z_{beam}=0$ and $z_{beam}=500$mm correspond to the two spherical mirrors $M_3$ and $M_4$ respectively and the positions $z_{beam}=1000$mm and $z_{beam}=1500$mm to the flat mirrors $M_1$ and $M_2$.**

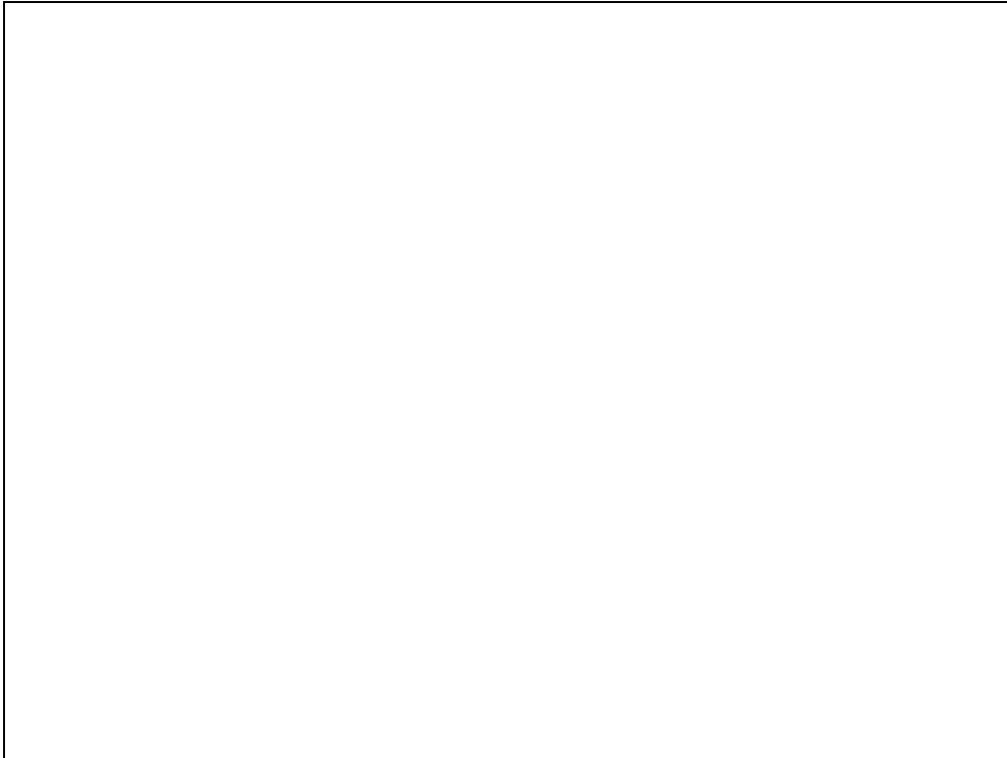

**Figure 15 Orientation of the elliptical profile of the fundamental mode of two tetrahedron cavities (see text) as a function of the unfolded coordinate along the mode propagation axis $z_{beam}$. The positions $z_{beam}=0$ and $z_{beam}=500$mm correspond to the two spherical mirrors $M_3$ and $M_4$ respectively.**